\documentstyle[prl,aps,multicol]{revtex}

\def\Lie{\pounds}                                  

\begin{document}

\title{Pseudotensors and quasilocal energy-momentum}

\author{Chia-Chen Chang \and
        James M. Nester\thanks{email: nester@joule.phy.ncu.edu.tw}}
\address{Department of Physics and Center for Complex Systems,
National Central University, Chungli, Taiwan 320, R.O.C.}
\author{Chiang-Mei Chen\thanks{email: chen@grg1.phys.msu.su}}
\address{Department of Theoretical Physics, Moscow State University,
         119899, Moscow, Russia}


\maketitle

\begin{abstract}
Early energy-momentum investigations for gravitating systems gave
reference frame dependent pseudotensors; later the quasilocal idea was
developed.  Quasilocal energy-momentum can be determined by the
Hamiltonian boundary term, which also identifies the variables to be
held fixed on the boundary.  We show that a pseudotensor corresponds
to a Hamiltonian boundary term.  Hence they are quasilocal
and acceptable; each is the energy-momentum density for a definite
physical situation with certain boundary conditions.  These conditions
are identified for well-known pseudotensors.
\end{abstract}

\pacs{PACS number(s):  4.20.Cv, 4.20.Fy}

\begin{multicols}{2} 

Energy-momentum can be regarded as the most fundamental conserved
quantity, being associated with a symmetry of the space-time geometry.
From Noether's theorem and translation invariance one obtains a
tensor, $T^{\mu\nu}$, which describes the density of energy-momentum
and defines a conserved energy-momentum as a consequence of its
satisfying the differential conservation law $\partial_\nu
T^{\mu\nu}=0$.  But it is not unique, since one can add an
arbitrary ``curl''---which shifts the zero of energy-momentum.  The
gravitational response, however, via Einstein's equation
$G^{\mu\nu}=\kappa T^{\mu\nu}$, detects the total energy-momentum
density for matter and interaction fields, and thereby removes the
classical uncertainty in the energy-momentum expressions---the
gravitational field is trivial only if the total energy
momentum vanishes.  Thus it seems ironic that the proper
identification of the contribution to the total energy-momentum from
the gravitational field itself has proved to be so elusive.  It was
natural to expect that, since these other sources exchange
energy-momentum with the gravitational field locally, gravity should
also have its own local energy-momentum density.

Attempts at identifying an energy-momentum density for gravity,
however, led only to various energy momentum complexes which are {\em
pseudotensors}, including those of Einstein \cite{Einstein},
Papapetrou \cite{Papa48}, Bergmann \cite{Berg53}, Landau and Lifshitz
\cite{LL62}, M{\o}ller \cite{Mol58}, and Weinberg \cite{Wein72}.
Pseudotensors are not covariant objects; they inherently depend on the
reference frame, and thus by their very nature cannot provide a truly
physical local gravitational energy-momentum density.  Indeed any such
quantity is precluded by the {\em equivalence principle} itself, since
a gravitational field should not be detectable at a point.
Consequently many have criticised the whole idea; e.g., an
influential textbook states:  {\em Anyone who looks for a magic
formula for ``local gravitational energy-momentum'' is looking for the
right answer to the wrong question.  Unhappily, enormous time and
effort were devoted in the past to trying to ``answer this question''
before investigators realized the futility of the enterprise.}
\cite{MTW} p.\ 467.  Hence the pseudotensor approach has been largely
abandoned (although interest continues, see e.g., \cite{DSV}).  A new
idea, {\em quasilocal} (i.e., associated with a closed 2-surface), was
proposed and has become widely accepted.  In view of the
role of gravity in identifying energy, this then becomes the most
fundamental notion of classical energy-momentum.  There have recently
been many quasilocal proposals \cite{quasilocal,Kij}.  Various
criteria have been advanced (see,
e.g., \cite{CY88}), in particular good limits to flat spacetime, weak
field, spatial infinity (ADM), and null infinity (Bondi).  However it
has now been recognized that there are an {\em infinite number} of
expressions satisfying these requirements \cite{Berq92}.  Clearly,
additional {\em principles} and {\em criteria} are needed.  The
Hamiltonian formalism includes such a principle, with the Hamiltonian
boundary term determining both the quasilocal quantities and the
boundary conditions \cite{Kij,CMChen}.  Here we show that this
Hamiltonian approach to quasilocal energy-momentum rehabilitates the
pseudotensors.

We identify energy with the value of the Hamiltonian.  The
Hamiltonian for a finite region,
 \begin{equation} H(N)=\int_{\Sigma}
N^\mu {\cal H}_\mu + \oint_{S=\partial \Sigma}{\cal B}(N),
\end{equation}
generates the spacetime displacement of a finite spacelike
hypersurface $\Sigma$ along a vector field $N^\mu$; it includes a
surface and a boundary term.  Noether's theorems guarantee that ${\cal
H}_\mu$ is proportional to the field equations.  Consequently the
value depends only on the boundary term ${\cal B}$, which gives the
{\em quasilocal energy-momentum}.  But the boundary term can be
modified.  (This is a particular case of the usual Noether conserved
current non-uniqueness.) \ Indeed it {\em is} necessary to adjust
$\cal B$ to give the correct asymptotic values \cite{RT74}.
Fortunately, ${\cal B}$ is not arbitrary.  A further {\em principle}
of the formalism controls its form:  {\em choose} the Hamiltonian
boundary term ${\cal B}$ so that the boundary term in $\delta H$
vanishes when the desired fields are held fixed on the boundary
(technically necessary for the variational derivatives to be well
defined).  Hence, we find a nice division:  the Hamiltonian density
${\cal H}_\mu$ determines the evolution and the constraint equations,
the boundary term ${\cal B}$ determines the boundary conditions and
the quasilocal energy-momentum.  There still remain many
possible boundary condition choices \cite{principle}.
Consequently there are various kinds of energy, each corresponding to
a different choice of boundary condition; this situation can be
compared with thermodynamics with its various
energies: internal, enthalpy, Gibbs, and Helmholtz.

For geometric gravity theories including general
relativity, the Hamiltonian can be succinctly obtained in terms of
differential forms \cite{coham}, which readily displays
the boundary term and its connection to pseudotensors.  Since we wish
to connect with traditional work we shall use a coordinate (holonomic)
basis $dx^\alpha$ and sometimes the (Hodge)
dual basis
$\eta^{\alpha\cdots}:=*(dx^\alpha\wedge\cdots)$.
The simplest analysis uses the connection along with
the metric as the dynamic variables.  The {\em curvature 2-form} is
given by
 $\Omega^\alpha{}_\beta := d\omega^\alpha{}_\beta+
\omega^\alpha{}_\gamma\wedge\omega^\gamma{}_\beta$,
where $\omega^\alpha{}_\beta=\Gamma^\alpha{}_{\beta\gamma}dx^\gamma$
is the {\em connection one-form}.  The Lagrangian density is the
Einstein-Hilbert scalar curvature 4-form
 ${\cal L}=\Omega^\alpha{}_\beta\wedge\eta_\alpha{}^\beta
=R\sqrt{-g}d^4x$.
The Hamiltonian 3-form can be constructed according to the pattern
$L=\dot q p-H$  by contracting the Lagrangian 4-form with the time
evolution vector field (dropping indices for brevity):
\begin{equation} i_N {\cal L}=i_N\Omega\wedge\eta+\Omega\wedge i_N\eta
=\Lie_N\omega\wedge\eta-{\cal H}(N);
\end{equation}
here the time derivative is given by the Lie derivative
(on form components $\Lie_N:=di_N+i_Nd$); the
Hamiltonian 3-form (without discarding total derivatives) is
\begin{equation}  {\cal H}(N)=
-\Omega\wedge i_N\eta-i_N\omega D\eta+d{\cal B}_M(N).
\label{Molham}
\end{equation}
The Hamiltonian density term includes
$N^\mu{\cal H}_\mu=-\Omega^\alpha{}_\beta\wedge
N^\mu\eta_\alpha{}^\beta{}_\mu=2N^\mu G^\nu{}_\mu\eta_\nu$,
a covariant expression which projects to the usual ADM Hamiltonian
density (see, e.g.,  Ch 21 in \cite{MTW}), along with a frame gauge
transformation generating term, $i_N\omega D\eta$, which vanishes
because the connection is symmetric and metric compatible.  When
integrated over a finite spatial hypersurface $\Sigma$, the value of
the Hamiltonian comes from the total differential term which (via the
generalized Stokes theorem) yields a boundary term with the 2-form
integrand
${\cal B}_M(N)=i_N\omega^\alpha{}_\beta\eta_\alpha{}^\beta=
N^\lambda{\cal M}_\lambda{}^{\mu\nu}(1/2)dS_{\mu\nu}$, where
$dS_{\mu\nu}:=(1/2)\epsilon_{\mu\nu\alpha\beta}dx^\alpha\wedge
dx^\beta$; the coefficients
\begin{equation}
{\cal M}_\lambda{}^{\alpha\sigma}
=2\sqrt{-g}g^{\beta[\sigma}\Gamma^{\alpha]}{}_{\beta\lambda}
=2\sqrt{-g}g^{\beta\sigma}g^{\alpha\mu}\partial_{[\beta}
g_{\mu]\lambda},
 \end{equation}
turn out to be the {\em superpotential} whose
divergence gives the M{\o}ller pseudotensor \cite{Mol58}.
(Note the extreme directness of this derivation of
M{\o}ller's expression.) \  The variation
of the Hamiltonian density (\ref{Molham}) yields (with $N$ fixed)
\begin{equation}
\delta{\cal H}_M=\hbox{field equation terms} +
 di_N(\delta\omega^\alpha{}_\beta\wedge\eta_\alpha{}^\beta),
\end{equation}
showing that the boundary condition implicit in (\ref{Molham}) is
of the Neumann type: the connection $\Gamma\sim\partial g$ is to be
held fixed. This calculation also reveals a serious deficiency: the
boundary term in the variation of the M{\o}ller Hamiltonian
will not vanish with the standard asymptotics: $\delta g\sim O(1/r)$,
$\delta \Gamma\sim O(1/r^2)$.

An almost obvious alternative (in view of the problem just mentioned)
is simply to replace the M{\o}ller Hamiltonian boundary term with
${\cal B}_E:=\omega^\alpha{}_\beta \wedge i_N \eta_\alpha{}^\beta
= N^\lambda{\cal U}_\lambda{}^{\mu\nu}(1/2)dS_{\mu\nu}$. Now the
coefficients
\begin{eqnarray}
{\cal U}_\lambda{}^{\mu\nu}
&=&
(-g)^{1/2}g^{\beta\sigma}\Gamma^\alpha{}_{\beta\gamma}
\delta^{\mu\nu\gamma}_{\alpha\sigma\lambda}
\nonumber \\&&\equiv
(-g)^{-\frac{1}{2}}g_{\lambda\tau}
\partial_\gamma
\left[-g(g^{\mu\tau} g^{\nu\gamma}-g^{\nu\tau}
g^{\mu\gamma}\right)],
\label{Freud}
\end{eqnarray}
are the {\em Freud} superpotential, whose divergence
gives the {\em Einstein} pseudotensor \cite{Einstein,Wal80}.
The Hamiltonian variation now contains a boundary term
of the form $di_N{\cal C}$ where
\begin{equation}
{\cal C}=-\omega^\alpha{}_\beta\wedge \delta \eta_\alpha{}^\beta=
-\Gamma^\alpha{}_{\beta\gamma}\delta(\sqrt{-g}g^{\beta\sigma})
\delta^{\nu\gamma}_{\alpha\sigma}(d^3x)_\nu,
\label{varFreud}
\end{equation}
showing that this choice of Hamiltonian boundary term is associated
with a Dirichlet type boundary condition: the contravariant
metric density is to be held fixed.

The possible forms of the Hamiltonian boundary
term have been considered in some detail elsewhere \cite{CMChen}.
There are various {\em choices} involved including:
%
 (1) the {\em representation} or {\em dynamic variables}, such as
the
metric, orthonormal frame, connection, spinors;
%
 (2) the {\em control mode}: such as a Dirichlet or
Neumann boundary condition for each dynamic variable;
 %
 (3) the {\em reference configuration}: this is given by the field
values that give vanishing energy-momentum (the standard choice is
Minkowski space).
%
 (4) the {\em displacement vector field} $N$:
which timelike displacement gives the energy?
which spatial displacement gives the linear
momentum?

We have shown how two famous pseudotensors naturally arise from
Hamiltonian boundary terms and how they are associated with field
boundary values.  In like manner we shall now explain how the
Hamiltonian boundary term approach to quasilocal energy-momentum
rehabilitates all of the other pseudotensors so they can be recognized
as legitimate.  There is a direct relationship between pseudotensors
and quasilocal expressions.  Every pseudotensor corresponds
to some {\em acceptable choice} of boundary expression.  Conversely,
every boundary expression defines a pseudotensor \cite{Chang}.

We consider the pseudotensor idea in some detail:
a suitable {\em superpotential}
 $H_\mu{}^{\nu\lambda}\equiv H_\mu{}^{[\nu\lambda]}$ is
selected and used to split the Einstein tensor thereby defining the
associated gravitational energy-momentum pseudotensor:
 \begin{equation}
\kappa \sqrt{-g} N^\mu t_\mu{}^\nu:=-N^\mu\sqrt{-g}G_\mu{}^\nu+
\frac{1}{2}\partial_\lambda( N^\mu H_\mu{}^{\nu\lambda}),
\end{equation}
where $\kappa=8\pi Gc^{-4}$ and we have inserted a vector field to
make the calculation more nearly covariant.  The usual
formulation is recovered by taking the components of the vector field
to be constant in the present reference frame; then Einstein's
equation, $G_\mu{}^\nu=\kappa T_\mu{}^\nu$,
can be rearranged into a form where the source is the {\em total}
effective energy-momentum pseudotensor
\begin{equation}
\partial_\lambda H_\mu{}^{\nu\lambda}=
2\kappa (-g)^{\frac{1}{2}}{\cal T}_\mu{}^\nu:=
2\kappa(-g)^{\frac{1}{2}}( t_\mu{}^\nu + T_\mu{}^\nu).
\end{equation}
An immediate consequence of the antisymmetry of the superpotential is
that ${\cal T}_\mu{}^\nu$ is a conserved
current: $\partial_\nu [(-g)^{1/2} {\cal T}_\mu{}^\nu]\equiv0$,
which integrates to give a conserved energy-momentum,
$N^\mu P_\mu:=\int  N^\mu {\cal T}_\mu{}^\nu (-g)^{1/2}(d^3x)_\nu$.
This should be contrasted with the covariant formula
\begin{equation}
\nabla_\nu T_\mu{}^\nu=\partial_\nu T_\mu{}^\nu-
\Gamma^\lambda{}_{\nu\mu}T_\lambda{}^\nu+
\Gamma^\nu{}_{\lambda\nu}T_\nu{}^\lambda=0,
\end{equation}
which {\em does not} lead to a conserved energy-momentum unless
$\Gamma=0$ (flat space).

A minor variation on the preceding analysis results from choosing a
superpotential with a contravariant index:
$H^{\mu\nu\lambda}\equiv H^{\mu[\nu\lambda]}$.
A further variation:
\begin{equation}
H^{\mu\nu\alpha}:=\partial_\beta H^{\mu\alpha\nu\beta},
\label{pattern}
\end{equation}
along with the symmetries
$H^{\mu\alpha\nu\beta}\equiv H^{\nu\beta\mu\alpha}\equiv
H^{[\mu\alpha][\nu\beta]}$ and $H^{\mu[\alpha\nu\beta]}\equiv0$,
leads to a {\em symmetric} pseudotensor---which then allows for a
simple definition of angular momentum, see \cite{MTW} \S20.2.  We can
cover these options simply by introducing a vector field, then, in the
following computations, we can easily make modifications like
$N^\mu H_\mu{}^{\nu\lambda}\longrightarrow N_\mu H^\mu{}^{\nu\lambda}$.

The energy-momentum within a finite region
\begin{eqnarray}
-P&&(N):=-
\int_{\Sigma} N^\mu {\cal T}_\mu{}^\nu \sqrt{-g}(d^3x)_\nu
\equiv\nonumber\\ &&
\int_{\Sigma} \bigl[ N^\mu \sqrt{-g}(\frac{1}{\kappa}
G_\mu{}^\nu-T_\mu{}^\nu) - \frac{1}{2\kappa}
\partial_\lambda (N^\mu H_\mu{}^{\nu\lambda})\bigr] (d^3x)_\nu
 \nonumber\\
&&\equiv\int_{\Sigma} N^\mu {\cal H}_\mu +
 \oint_{S=\partial \Sigma}{\cal B}(N)\equiv H(N),
\label{Ham}
\end{eqnarray}
is seen to be just the value of the Hamiltonian.  Note that
${\cal H}_\mu$ is the covariant form of the ADM Hamiltonian density,
which has a vanishing numerical value, so that
the value of the Hamiltonian is determined purely by the boundary term
${\cal B}(N)=-N^\mu (1/2\kappa)H_\mu{}^{\nu\lambda}
(1/2)dS_{\nu\lambda}$.
Thus for any pseudotensor the associated {\em superpotential} is
naturally a Hamiltonian boundary term.  Moreover the energy-momentum
defined by such a pseudotensor does not really depend on the local
value of the reference frame, it is actually {\em quasilocal}---it
depends (through the superpotential) on the values of the
reference frame (and the fields) only on the boundary of a region.

Even more important, the Hamiltonian approach endows these values with
a physical significance.  To understand the {\em physical meaning} of
the quasilocalization, calculate the boundary term in the Hamiltonian
variation:
\begin{equation}
-{1\over4\kappa}\left[\delta\Gamma^\alpha{}_{\beta\lambda}N^\mu
\sqrt{-g}g^{\beta\sigma}\delta^{\tau\rho\lambda}_{\alpha\sigma\mu}
+\delta(N^\mu H_\mu{}^{\tau\rho})\right]dS_{\tau\rho}.
\end{equation}
For the {\em Einstein} pseudotensor, we use the Freud superpotential
(\ref{Freud}) as
the Hamiltonian boundary term in (\ref{Ham}).  Then the boundary term
in the Hamiltonian variation \cite{iNwDelta} has the integrand
$\delta(\sqrt{-g} g^{\beta\sigma}N^\mu) \Gamma^\alpha{}_{\beta\lambda}
\delta^{\tau\rho\lambda}_{\alpha\sigma\mu}$,
which shows not only that  $\sqrt{-g}g^{\beta\sigma}$
is to be held fixed on the boundary, but also that the appropriate
displacement vector field is $N^\mu=$ constant, and the reference
configuration here (as well as in the other cases below) is Minkowski
space with a Cartesian reference frame.

This calculation is easily adapted to some other cases just by
adjusting $N^\mu$.  The {\em Bergmann} pseudotensor \cite{Berg53},
given by
 $2\kappa\sqrt{-g}{\cal T}^{\mu\nu}_{B}
:=\partial_\lambda (g^{\mu\gamma} {\cal U}_\gamma{}^{\nu\lambda})$,
leads to the Hamiltonian boundary variation term
$\delta(\sqrt{-g} g^{\beta\sigma} g^{\mu\nu}N_\mu)
\Gamma^\alpha{}_{\beta\lambda}
\delta^{\tau\rho\lambda}_{\alpha\sigma\nu} dS_{\tau\rho}$,
revealing that we are to fix
$\sqrt{-g}g^{\beta[\sigma}g^{\nu]\mu}$
on the boundary and use the displacement (co)vector
 $N_\mu=N^0_\mu=\mbox{constant}$.  This last statement takes
on a more proper geometric form in terms of an auxiliary (background)
metric ${\bar g}{}_{\mu\nu}$ having the Minkowski values in this
coordinate system, then the desired displacement {\em vector} is
$N^\alpha={\bar g}{}^{\alpha\gamma}N^0_\gamma$.

The {\em Landau-Lifshitz} pseudotensor \cite{LL62} is slightly more
complicated, being given by a weighted density
 $2\kappa(-g){\cal T}^{\mu\nu}_{LL}
:=\partial_\lambda (\sqrt{-g}g^{\mu\gamma}
 {\cal U}_\gamma{}^{\nu\lambda})$.
The easiest way to handle this is to introduce, where necessary to
obtain the proper geometric density weights, ``extra'' factors
of $(-\bar g)^{1/2}$, the Jacobian factor
for the flat metric (numerically constant in Cartesian coordinates).
This leads to the conclusion that the displacement vector should be
$N^\mu=g^{\mu\nu}(g/{\bar g})^{1/2}N^0_\nu$ and the quantity
to be held fixed on the boundary is $(-g)g^{\beta[\sigma}g^{\nu]\mu}$.

The three pseudotensors just discussed are associated with similar but
distinct Dirichlet type boundary conditions which are algebraic in
terms of the metric.  On the other hand, the M{\o}ller pseudotensor
has a simple Neumann type condition.  While the detailed physical
significance of these conditions has not yet been probed, such an
investigation seems straightforward.  In contrast, the remaining
pseudotensors in our survey are associated with more complicated
boundary conditions.

In the context of Eq.~(\ref{pattern}), Goldberg \cite{Gold58}
discussed the general form
$H^{\mu\alpha\nu\beta}\equiv
H^{\mu\nu}H^{\alpha\beta}-H^{\alpha\nu}H^{\mu\beta}$
 including various weighted densities.  Because of the symmetries,
the associated pseudotensor,
$\sqrt{-g}{\cal T}^{\mu\nu}(H):=\partial_\alpha\partial_\beta
 H^{\mu\alpha\nu\beta}$,
is guaranteed to be symmetric and conserved for all symmetric
$H^{\mu\nu}$.
This pattern can nicely accommodate the Landau-Lifshitz version with
$H^{\mu\nu}:=\sqrt{-g}g^{\mu\nu}$.
More generally, we note that
 ${\cal T}(H_1+H_2)-{\cal T}(H_1)-{\cal T}(H_2)$
is also identically conserved leading to the more general pattern
$H^{\mu\alpha\nu\beta}=
H_1^{\mu\nu}H_2^{\alpha\beta}-H_1^{\mu\beta}H_2^{\nu\alpha}
+H_2^{\mu\nu}H_1^{\alpha\beta}-H_2^{\mu\beta}H_1^{\nu\alpha}$.
With $H_2^{\alpha\beta}={\bar g}{}^{\alpha\beta}$
 we can now accommodate the pseudotensors of Papapetrou \cite{Papa48}
($H_1^{\mu\nu}=\sqrt{-g}g^{\mu\nu}$)
and Weinberg
\cite{Wein72}
($H_1^{\mu\nu}=-h^{\mu\nu}+
\frac{1}{2}{\bar g}{}^{\mu\nu}h^\lambda{}_\lambda$,
where
$h_{\mu\nu}:=g_{\mu\nu}-{\bar g}{}_{\mu\nu}$ and indices are
raised with ${\bar g}{}^{\mu\nu}$).  The Hamiltonian variations then
lead to boundary conditions involving rather complicated combinations
of $\delta\Gamma^\alpha{}_{\mu\nu}$ and $\delta(\sqrt{-g}g^{\mu\nu})$
or $\delta h_{\mu\nu}$, respectively.

In summary, because of the very nature of the gravitational
interaction and its elusive contribution, the localization of total
energy-momentum has remained an outstanding fundamental puzzle.  The
earlier pseudotensor approach was considered to be unsatisfactory.  A
newer idea is quasilocal.  Quasilocal energy-momentum can be obtained
from the Hamiltonian.  For a finite region it includes a boundary term
which plays the key role, determining both the boundary conditions and
the quasilocal values.  Consequently there are (as in
thermodynamics) many different physical kinds of energy, each
corresponding to a different boundary condition.  We have shown that
every energy-momentum pseudotensor is associated with a legitimate
Hamiltonian boundary term.  Hence the pseudotensors are quasilocal and
acceptable.  Each is the energy-momentum density for some definite
physical situation.  Via the Hamiltonian approach one can identify the
necessary boundary conditions and thereby appreciate the physical
significance of the associated energy-momentum quasilocalization.

Our analysis reclaims the pseudotensor work of the past to
its rightful place:  concerning a special class of quasilocal
energy-momentum.  Moreover it is additional evidence that the
Hamiltonian boundary term approach to energy provides an effective
ordering principle for the various quasilocal energy-momentum
expressions.


This work was supported by the National Science Council
of the R.O.C. under grants NSC88-2112-M-008-018,
 NSC87-2112-M-008-007,
 NSC86-2112-M-008-009.

\end{multicols} 


\begin{references}

\bibitem{Einstein}
        See, e.g.,
           A. Trautman,
        in {\sl Gravitation: an Introduction to Current Research},
         ed. L. Witten (Wiley, New York, 1962), 169-198.

\bibitem{Papa48}
     A. Papapetrou, {Proc. Roy. Irish Acad. A \bf52}, 11-23 (1948);
%
     S. N. Gupta, {Phys. Rev. \bf 96}, 1683-1685 (1954);
%
     this pseudotensor has more recently been rediscovered by
     D. Bak, D. Cangemi, and R. Jackiw,
     {Phys. Rev. D \bf 49}, 5173-5181 (1994).

\bibitem{Berg53} P. G. Bergmann and R. Thompson,
                 {Phys. Rev. \bf 89}, 400-407 (1953).

\bibitem{LL62}
    L. D. Landau and E. M. Lifshitz,
    {\sl The Classical Theory of Fields}, 2nd ed.
    (Reading, Mass.: Addison-Wesley, 1962).

\bibitem{Mol58} C. M{\o}ller,
                 {Ann. Phys. \bf 4}, 347-371 (1958).

\bibitem{Wein72} S. Weinberg,
              {\sl Gravitation and Cosmology}
               (Wiley, New York, 1972);
              the same energy-momentum density is used in
              \cite{MTW}, \S 20.3.

\bibitem{MTW} C. W. Misner, K. Thorne, and J. A. Wheeler,
       {\sl Gravitation} (Freeman, San Francisco, 1973).

\bibitem{DSV}
   M. Dubois-Violette and J. Madore,
                {Comm. Math. Phys. \bf 108}, 213-223 (1987);
%
   J. Frauendiener,
               {Class. Quant. Grav. \bf 6}, L237-241 (1989);
%
   L. B. Szabados,
               {Class. Quant. Grav. \bf 9}, 2521-2541 (1992);
%
   J. M. Aguirregabiria, A. Chamorro, and K. S. Virbhadra,
               {Gen. Rel. Grav. \bf 28}, 1393-1400 (1996).

\bibitem{quasilocal}
    See
    J. D. Brown and J. W. York, Jr.,
    {Phys. Rev. D \bf 47}, 1407-1419 (1993) and the references cited
    therein.
 %
        Some more recent works are:
%
    S. Lau,
    {Class. Quantum Grav. \bf 10}, 2379-2399 (1993);
%
    L. B. Szabados,
    {Class. Quant. Grav \bf 11}, 1847-1866 (1994);
%
    S. A. Hayward,
    {Phys. Rev. D \bf 49}, 831-839 (1994);
%
 J. Katz, J. Bi{\v c}\'ak and D. Lynden-Bell,
               {Phys. Rev. D \bf 55}, 5957-5969 (1997).

\bibitem{Kij}
    J. Jezierski and J. Kijowski,
    {Gen. Rel. Grav. \bf 22}, 1283-1307 (1990).
%
    J. Kijowski,
    {Gen. Relativ. Grav. \bf 29}, 307-343 (1996).

\bibitem{CY88}
    D. Christodoulou and S. T. Yau,
    {Some Remarks on the Quasi-local Mass} in
    {\sl Contemporary Mathematics \bf 71}:
   {\sl Mathematics and General Relativity},
    ed. J. Isenberg
    (Providence RI: AMS, 1988), pp. 9-16.

\bibitem{Berq92}
    G. Bergqvist,
    {Class. Quantum Grav. \bf 9}, 1917-1922 (1992).

\bibitem{CMChen}
    C. M. Chen, J. M. Nester, and R. S. Tung,
    {Phys. Lett. \bf 203A}, 5-11 (1995).
%
    C. M. Chen and J. M. Nester,
    {\sl Quasilocal quantities for GR and other gravity theories}
     Class. Quant. Grav., to be published, {\tt gr-qc/9809020}.

\bibitem{RT74}
    T. Regge and C. Teitelboim,
    {Ann. Phys. \bf 88}, 286-319 (1974).

\bibitem{principle}
Some further principle is needed.  We have advocated {\em covariance}.
For each dynamical field it was found \cite{CMChen},
using {\em
symplectic} techniques \cite{KT79}, that there are only two choices
for $\cal B$ corresponding to holding a {\em covariant} quantity fixed
on the boundary (essentially {\em Dirichlet \rm or \em Neumann}
``control mode'').  The pseudotensor boundary terms discussed here do
not satisfy this covariance principle.

\bibitem{KT79}
    J. Kijowski and W. M. Tulczyjew,
    {\sl A Symplectic Framework for Field Theories}
    (Berlin: Springer Verlag, 1979), Lecture Notes in Physics
    {\bf 107}.

\bibitem{coham}
    J. M. Nester,
    {The Gravitational Hamiltonian}, in
    {\sl Asymptotic Behavior of Mass and Space-Time Geometry},
    ed. F. Flaherty
    (Berlin: Springer-Verlag)
    Lecture Notes in Physics {\bf 202}, 155-163 (1984);
%
    {Mod. Phys. Lett. \bf A6}, 2655-2661  (1991).

\bibitem{Wal80} For a nice discussion of the Einstein pseudotensor and
      the Freud superpotential using differential forms see R.  P.
      Wallner, {Acta Physica Austriaca \bf 52}, 121-124 (1980).

\bibitem{Chang} C. C. Chang,
      {\sl Gravitational Energy-Momentum: Quasilocal and Pseudotensors}
       MSc. Thesis, National Central Univ., Chungli 1998, (unpublished).

\bibitem{iNwDelta} This differs slightly from Eq.~(\ref{varFreud})
because we have not included the $i_N\omega D\eta$ term of
Eq.~(\ref{Molham}) in Eq.~(\ref{Ham}).

\bibitem{Gold58} J. N. Goldberg,
               {Phys. Rev. \bf 111}, 315-320 (1958).

\end{references}
\end{document}